\documentclass[twocolumn,showpacs,secnumarabic,tightenlines,amssymb,
amsmath,nobibnotes,aps,prb]{revtex4}
\usepackage{graphicx}
\usepackage{dcolumn}
\usepackage{bm}
\begin{document}
\preprint{APS/123-QED}
\title{Nonequilibrium Kubo Formula 
of Finite Conductor Connected to Reservoirs based on Keldysh Formalism}
\author{Tatsuya Fujii}
\affiliation{Institute for Solid State Physics, 
University of Tokyo, Kashiwa 277-8581, Japan}
%
\date{\today}
\begin{abstract}
We show that the density matrix 
for a finite conductor attached to reservoirs
obtained by Keldysh formalism 
is of MacLennan-Zubarev form. 
On the basis of the fact that 
the density matrix is the invariant part 
proposed by Zubarev, 
it is shown that Keldysh formalism 
may describe the irreversible processes and 
steady-state feature of the system. 
An important consequence of 
the MacLennan-Zubarev form 
of the density matrix is 
a generalization of the Kubo formula 
in a nonequilibrium case. 
On the basis of the result, we propose the formula of 
shot noise and a nonequilibrium identity 
between differential conductance, noise power and 
shot noise as a generalized Nyquist-Johnson relation. 
\end{abstract}
\pacs{73.63.Kv}
\maketitle
\section{Introduction}
Recently many-body problems in nonequilibrium 
steady states have 
been studied in mesoscopic devices. 
Keldysh formalism 
\cite{Keldysh,langreth,rammer,kamenev1,kamenev2} 
has been applied to study 
various systems. 
It has been used successfully 
to calculate observable quantities 
such as differential conductance and noise power. 

However, in mesoscopic systems with electron 
correlation, progress has been limited even 
in the basic understanding of nonequilibrium steady states.
Of course the linear response theory is well established. 
In mesoscopic systems, 
the Kubo formula leads to the Nyquist-Johnson relation 
between conductance and noise power. 
However, there has been little discussion on 
the generalization of the Kubo formula and 
the relation between physical quantities in 
many-body systems in a nonequilibrium situation. 

One possible means of discussing general features 
under the nonequilibrium condition 
independent of different systems may be to consider 
density matrices. 
In this study, we consider a conductor in series with 
equilibrium reservoirs with a finite bias voltage. 
There is an approach to formulating 
the density matrix of this system 
\cite{Hershfield2,Hershfield3}, 
in which the density matrix is given by Keldysh formalism, 
although the term Keldysh formalism is not used. 
The obtained density matrix is 
similar to the equilibrium ensemble. 
The density matrix is expressed 
as an infinite series of operators, 
which are generated by the perturbative expansion 
of the s-matrix. Therefore, the density 
matrix has remained to be a formal expression. 
The generalization of the Kubo formula under a 
nonequilibrium condition has not been 
successful. 

On the other hand, 
there has been progress by a different method. 
By using a rigorous approach of the ${\rm C}^{*}$ algebra, 
a density matrix of 
MacLennan-Zubarev form has been obtained 
for a finite system attached to reservoirs 
\cite{stasaki1,stasaki2}. 
Concerning the transport properties of a 
one-dimensional noninteracting lattice or 
a quantum dot using the slave-boson technique, 
it has been reported that 
a current-current correlation function is given by 
the sum of differential conductance and 
an additional term. 
However, the physical picture and origin of 
the additional term have not been clarified
\cite{stasaki1,stasaki3}. 

In this study we show that 
the Keldysh formalism of the density matrix 
of a finite conductor attached to reservoirs 
is of MacLennan-Zubarev form. 
We show that Keldysh formalism
has sufficient features to describe 
the nonequilibrium steady state. 
We finally find that the MacLennan-Zubarev form of the density matrix 
leads to a nonequilibrium Kubo formula. 
On the basis of this form, we propose a $\it nonequilibrium {\ }identity$ 
between differential conductance, noise power and 
shot noise, which is an extended Nyquist-Johnson relation 
in correlated systems in a nonequilibrium case. 
\section{System and Hamiltonian}
We discuss a 
finite conductor attached to the left and right 
infinite reservoirs through boundary couplings. 
The system Hamiltonian is defined as 
\begin{eqnarray}
H \equiv H_{c} + H_{L} + H_{R} + H_{cL,R}, 
\label{hamiltonian0}
\end{eqnarray}
where $H_{c},H_{L,R}$ and $H_{cL,R}$ describe 
the conductor, left and right reservoirs, and 
boundary couplings, respectively. 
The conductor 
part of the Hamiltonian is given by 
the sum of the noninteracting and 
interacting parts, $H_c \equiv H_{c0}+H_{c1}$. 
We stress that, in the following analysis, the correlation 
effect is included. 
Generally, in the preparation of the Keldysh Green function, 
the system Hamiltonian $H$ is 
divided into the nonperturbative and perturbative terms
\cite{Caroli,Hershfield,Meir} as 
\begin{eqnarray}
H=H_0 + H_1. 
\label{hamiltonian}
\end{eqnarray}
The nonperturbative term $H_0 \equiv H_{c0}+H_{L}+H_{R}$ 
describes 
three independent systems, which are the noninteracting 
conductor and infinite reservoirs. 
The perturbative term $H_1 \equiv H_{cL,R}+H_{c1}$ 
is the sum of boundary couplings and the interaction 
inside the conductor. 
In many cases, $H_{c1}$ is treated using a perturbation theory 
to calculate the Green function in Keldysh formalism. 
Notice that 
the following results do not depend on 
the method used to divide $H$ into $H_0$ and $H_1$, 
which will be shown explicitly in section 4. 

Clearly, the energy scale of $H_1$ is much smaller 
than that of $H_0$ because $H_1$ 
describes the effects in the finite conductor. 
Thus, it may be expected that 
the expansion of the s-matrix is a well-behaved series. 

\section{
Adiabatic Switching-on of $H_1$ in Keldysh Formalism
}
Let us begin with the Keldysh formalism 
of a finite conductor connected to 
infinite reservoirs\cite{Caroli,Hershfield,Meir}. 
Generally, the discussion based on Keldysh formalism 
is concentrated on the Keldysh contour of the s-matrix and 
the Keldysh Green function. 
We carefully analyze 
the time dependence of Keldysh 
formalism in which the perturbative term is 
imposed adiabatically. 

In the initial state at $t_0=-\infty$, the conductor and 
infinite reservoirs are decoupled in equilibrium state, 
where 
the initial density matrix is supposed to be given by 
$\rho_{0} \equiv e^{-\beta (H_0
-\mu N_c -\mu_{L} N_L -\mu_{R} N_R -\Omega _0)}$ and 
$\Omega _0$ is the initial thermodynamic potential. 
The reservoirs are tuned to different chemical potentials. 
For simplicity, we define 
$(\mu _L +\mu _R )/2 \equiv \mu \equiv 0$ and 
$(\mu _L -\mu _R )/2 \equiv eV/2$. 
We can rewrite the initial density matrix as 
\begin{eqnarray}
\rho _{0} = e^{-\beta (H_0 -eV/2 (N_L -N_R) -\Omega _0)}. 
\label{inidens}
\end{eqnarray}
In the following, we assume 
$[H_0, N_L -N_R ]=0$. 

The perturbative term 
is turned on adiabatically as $g e^{-\epsilon |t|} H_1$, 
where $g$ is introduced to tune the magnitude of 
the coupling constant, 
and $\epsilon$ is a positive infinitesimal number. 
We define the total Hamiltonian as 
\begin{eqnarray}
H_{\epsilon} \equiv H_0 + g e^{-\epsilon |t|} H_1.
\label{hamiltonian2}
\end{eqnarray}
The time evolution of the system obeys 
the Neumann equation 
\begin{eqnarray}
{\rm i} \frac{\partial}{\partial t} \rho _{\epsilon}(t)
= [H_{\epsilon}, \rho _{\epsilon}(t)], 
\label{neumann}
\end{eqnarray}
where the initial state is given by eq. (\ref{inidens}), 
and $[A,B] \equiv AB-BA$ is the commutation relation. 
The formal solution of the Neumann equation 
is given by 
\begin{eqnarray}
\nonumber 
&&\rho _{\epsilon}(t)
= U_{\epsilon}(t,t_0) \rho _{\epsilon}(t_0) U_{\epsilon}(t_0,t) \\
\nonumber
&&{\qquad }
= e^{-{\rm i}H_0 t} S_{\epsilon}(t,t_0) e^{{\rm i}H_0 t_0}
\rho _{\epsilon}(t_0) 
e^{-{\rm i}H_0 t_0} S_{\epsilon}(t_0,t) e^{{\rm i}H_0 t}. \\
\label{formalsol}
\end{eqnarray}
$U_{\epsilon}(t,t_0)$ 
or its equivalent $S_{\epsilon}(t,t_0)$ satisfies 
\begin{eqnarray}
&&{\!\!\!\!\!\!}
{\rm i} \frac{\partial}{\partial t}U_{\epsilon}(t,t_0) 
= H_{\epsilon} U_{\epsilon}(t,t_0), \\
&&{\!\!\!\!\!\!}
{\rm i} \frac{\partial}{\partial t}S_{\epsilon}(t,t_0) 
= g e^{-\epsilon |t|} H_1 (t) S_{\epsilon}(t,t_0), 
\label{equations}
\end{eqnarray}
where $U_{\epsilon}(t,t_0)= 
e^{-{\rm i}H_0 t} S_{\epsilon}(t,t_0) e^{{\rm i}H_0 t_0}$. 
In the following analysis, we employ 
the explicit forms $U_{\epsilon}(t,0)$ and 
$S_{\epsilon}(t,0)$. 
Although the expressions 
of $U_{\epsilon}(t,t_0)$ and 
$S_{\epsilon}(t,t_0)$ are well known, 
here we show them explicitly for later use as 
\begin{eqnarray}
\nonumber
& &{\!\!\!\!\!\!\!\!\!\!}
U_{\epsilon}(t,t_0) = 1+ \sum^{\infty}_{n=1} 
\frac{(-{\rm i})^n}{n!} \int^{t}_{t_0} {\!} {\rm d}t_1 \cdots {\rm d}t_n 
%
T [H_{\epsilon 1} \cdots H_{\epsilon n}], \\
\nonumber
& &{\!\!\!\!\!\!\!\!\!\!}
S_{\epsilon}(t,t_0) = 1+ \sum^{\infty}_{n=1} 
\frac{(-{\rm i}g)^n}{n!} \int^{t}_{t_0} {\!} {\rm d}t_1 \cdots {\rm d}t_n \\
\nonumber
& &{\qquad \qquad \qquad {\ }\,\,}
e^{-\epsilon (|t_1|+ \cdots |t_n|)} 
T [H_1(t_1) \cdots H_1(t_n)], \\
\label{smatrix}
\end{eqnarray}
where $T$ is 
the time-ordering operator 
and 
$H_{\epsilon n} \equiv H_0 + ge^{-\epsilon |t_n|}H_1$.

The expectation value of any operator $\cal O$ is given by 
\begin{eqnarray}
{\rm Tr} \{{\cal O} \rho _{\epsilon} (t) \}
= {\rm Tr} \{{\cal O}_H (t) \bar{\rho}_{\epsilon} \}
\equiv \langle {\cal O}_H (t) \rangle  , 
\label{expectation}
\end{eqnarray}
where $\bar{\rho}_{\epsilon} \equiv \rho _{\epsilon} (0)$. 
The Heisenberg representation is 
defined by 
\begin{eqnarray}
\nonumber
&&
{\cal O}_H (t) = 
U_{\epsilon}(0,t) {\cal O} U_{\epsilon}(t,0) =
S_{\epsilon}(0,t) {\cal O}(t) S_{\epsilon}(t,0), \\
\label{heisenberg}
\end{eqnarray}
where 
${\cal O}(t) = e^{{\rm i}H_0 t} {\cal O} e^{-{\rm i}H_0 t}$ 
is the interaction representation. 
The physical quantities are defined 
in the limit $g \rightarrow 1, \epsilon \rightarrow 0 $. 
On the basis of the observation that 
$\displaystyle \lim_{g \rightarrow 1, \epsilon \rightarrow 0} 
H_{\epsilon} = H$. In Keldysh formalism, 
$\cal O$ 
is supposed to exist in the limit 
$g \rightarrow 1, \epsilon \rightarrow 0$. 
\section{MacLennan-Zubarev Form}

Let us focus on the density matrix $\bar{\rho}_{\epsilon}$ 
defined by 
$\bar{\rho}_{\epsilon}
= S_{\epsilon}(0,-\infty) \rho _{0} S_{\epsilon}(-\infty ,0),
$
which may be rewritten as 
\begin{eqnarray}
\nonumber
&&{\!\!\!\!\!\!\!\!\!\!\!}
\bar{\rho}_{\epsilon}
= {\rm exp} \{ -\beta S_{\epsilon}(0,-\infty)( H_0
-eV/2 (N_L -N_R) \\
&&{\qquad \qquad \qquad \qquad \qquad } 
-\Omega _0) S_{\epsilon}(-\infty ,0) \}. 
\label{formal}
\end{eqnarray}
This is simply a formal expression. 
In standard calculations of Keldysh formalism, 
the s-matrix is almost always perturbatively expanded. 
In practice, we can sometimes estimate observable quantities 
based on some approximations 
even if we do not have the complete knowledge of 
the density matrix. 
In this paper, we try to analyze the structure 
of the density matrix, which leads to a deeper understanding 
of the nonequilibrium properties of mesoscopic systems. 

Here, in order to study $\bar{\rho}_{\epsilon}$, 
we derive a key relation for 
\begin{eqnarray}
{\!\!\!\!\!\!}
\bar{A}_{\epsilon} \equiv S_{\epsilon}(0,-\infty) A 
S_{\epsilon}(-\infty,0), 
\label{abar}
\end{eqnarray}
where the operator $A$ satisfies $[A,H_0]=0$. 
We notice concerning $A=H_0$ that a useful expression of 
$\bar{A}_{\epsilon}$ is 
\begin{eqnarray}
\nonumber
&&S_{\epsilon}(0,-\infty) H_0 S_{\epsilon}(-\infty,0) \\
&&{\ } 
=H_0 + gH_1 -{\rm i} \epsilon g
\frac{\partial S_{\epsilon}(0,-\infty)}{\partial g}
S_{\epsilon}(-\infty,0), 
\label{gellman}
\end{eqnarray}
which is obtained using
the Gell-Mann and Low theorem\cite{fetter,nozierestxt}. 
In the following, we derive a 
different expression for the arbitrary 
operator $A$ that satisfies $[A,H_0]=0$. 

First, we discuss the commutation relation between 
$A$ and the s-matrix. 
\begin{widetext}
\begin{eqnarray}
\nonumber
&&[A,S_{\epsilon}(0,-\infty)] =
\sum^{\infty}_{n=1} 
\frac{(-{\rm i}g)^n }{n!} \int^{0}_{-\infty} {\rm d}t_1 \cdots {\rm d}t_n 
e^{-\epsilon (|t_1|+ \cdots +|t_n|)} 
({\rm i}g)^{-1} 
n \sum_{l=1}^{n} 
{}_{n-1}C_{l-1} \\
&&
\nonumber
{\quad }
T [H_1(t_1) \cdots H_1(t_{l-1})] J_A(t_l)
T [H_1(t_{l+1}) \cdots H_1(t_{n})], 
{\ } 
\{t_1, \cdots ,t_{l-1} \} > t_l > \{t_{l+1}, \cdots ,t_{n} \}. \\
%
%
%
\label{commutation}
\end{eqnarray}
\end{widetext}
The "current" corresponding to 
the operator $A$ is defined as 
\begin{eqnarray}
J_{AH}(t) \equiv -\frac{\partial}{\partial t} A_H (t).
\label{current}
\end{eqnarray}
The calculation of eq.(\ref{commutation}) can be 
performed as follows: 
We expand the s-matrix and consider the $n$th 
term in the series. 
Among the 
$n!$ possible time orderings, 
a certain term of $t_p > t_q >  \cdots > t_r$ 
is chosen. 
The commutation relation 
$[A,H_1(t_p) H_1(t_q) \cdots H_1(t_{r})]$ 
may be rewritten using $[A,H_1(t)]=({\rm i}g)^{-1} J_A (t)$ 
for $[A,H_0]=0$. 
With reference to the time 
variable $t_l$ of $J_A$, we can categorize times 
larger $\{t_1, \cdots ,t_{l-1} \}$ and 
smaller $\{t_{l+1}, \cdots ,t_{n} \}$ than $t_l$. 
We sum possible time orderings except $J_A (t_l)$ 
while taking advantage of dummy integration variables. 
Finally, we obtain 
\begin{eqnarray}
{\quad}
[A,S_{\epsilon}(0,-\infty)] =
-\int _{-\infty}^{0} {\rm d}t e^{-\epsilon |t|} 
J_{AH}(t) S_{\epsilon} (0,-\infty). 
\label{commutationas}
\end{eqnarray}
It is easy to rewrite this as 
\begin{eqnarray}
\nonumber
&&{\!\!\!\!\!\!\!\!\!\!}\bar{A}_{\epsilon}
=S_{\epsilon}(0,-\infty) A S_{\epsilon}(-\infty ,0) \\
&&
{\!\!\!}=A+ \int _{-\infty}^{0} {\rm d}t e^{-\epsilon |t|} J_{AH}(t). 
\label{adiabatic}
\end{eqnarray}
Integration by parts in eq.(\ref{adiabatic}) also gives, 
\begin{eqnarray}
{\quad}
\bar{A}_{\epsilon}
= \epsilon \int _{-\infty}^{0} {\rm d}t e^{-\epsilon |t|} A_{H}(t). 
\label{adiabatic2}
\end{eqnarray}

Note that 
$\bar{A}_{\epsilon}$ is the invariant part 
of the operator $A$ defined by Zubarev\cite{zubarevtxt}.
The concept of the invariant part will play 
an essential role in the next section. 

By using eq. (\ref{adiabatic}), the density matrix 
of eq. (\ref{formal}) can be evaluated as 
\begin{eqnarray}
\nonumber
&&{\!\!\!\!\!\!\!\!\!}
\bar{\rho}_{\epsilon} 
={\rm exp} \left\{ -\beta(H_0 + \int ^{0}_{-\infty}
{\rm d}t e^{-\epsilon |t|}  J_{eH}(t) \right. \\
&&{\!\!\!}- \left. eV/2(N_L -N_R) -V \int ^{0}_{-\infty}
{\rm d}t e^{-\epsilon |t|}  J_{cH}(t) -\Omega _0) \right \}, 
\nonumber \\
\label{zubarev}
\end{eqnarray}
where $J_{eH}(t)$ and $J_{cH}(t)$ are 
the energy change 
and charge current, respectively, which 
are obtained by setting $A=H_0$ and $A=e/2(N_L -N_R)$ 
in eq. (\ref{current}). 

The obtained density matrix is nothing but a type of 
MacLennan-Zubarev 
form of the density matrix 
\cite{maclennan,zubarevtxt}, 
which is characterized by 
the time-integral of 
the "current" 
on the negative semi-infinite interval. 
We have obtained this form for the density matrix 
by dividing the system into 
$H_0=H_{c0}+H_L +H_R$ and $H_1=H_{cL,R}+H_{c1}$. 
On the other hand, for the case of 
$H_0=H_L +H_R$ and $H_1=H_{cL,R}+H_{c1}+H_{c0}$, 
it was obtained using a rigorous approach 
of a ${\rm C}^{*}$ algebra\cite{stasaki2}. 
Here, let us show that 
the obtained density matrix does not depend on 
the method used to divide the Hamiltonian. 

In eq. (\ref{adiabatic}), 
\begin{eqnarray}
{\ }
S_{\epsilon}(0,-\infty) H_0 S_{\epsilon}(-\infty,0) 
=H_0 + \int _{-\infty}^{0} {\rm d}t e^{-\epsilon |t|} J_{eH}(t), 
\label{adiabaticH0}
\end{eqnarray}
seems to be dependent on the method used to divide $H$ 
into $H_0$ and $H_1$. 

As pointed out previously, 
$S_{\epsilon}(0,-\infty) H_0 S_{\epsilon}(-\infty ,0)$ 
is rewriten as eq.(\ref{gellman}). 
On the right-hand side of eq. (\ref{gellman}), 
$H_0 +gH_1$ is clearly independent of the method used to 
divide $H$ into $H_0$ and $H_1$ in the limit 
$g \rightarrow 1,\epsilon \rightarrow 0$; 
hence, we discuss the last term 
\begin{eqnarray}
\Delta \Omega_{\epsilon} \equiv {\rm i} \epsilon g
\frac{\partial S_{\epsilon}(0,-\infty)}{\partial g}
S_{\epsilon}(-\infty,0).
\label{delomega}
\end{eqnarray}
%
Using $S_{\epsilon}(0,-\infty)=U_{\epsilon}(0,-\infty)$ 
and $U_{\epsilon}(0,-\infty)=U_{\epsilon}(-\infty,0)$, 
we obtain 
\begin{eqnarray}
\Delta \Omega_{\epsilon} = \epsilon 
\sum_{n=1}^{\infty} \frac{(-{\rm i}g)^{n-1}}{n!}
\int^{\infty}_{-\infty} {\rm d}t_1 \cdots {\rm d}t_n 
T[H_{\epsilon 1} \cdots H_{\epsilon n}].
\label{delomega2}
\end{eqnarray}
Now, we assume that $H$ is given by 
the sum of the operators $A$, $B$ and $C$ as $H=A+B+C$. 
We compare two cases: $H_0 =A+B,H_1=C$ and 
$H'_0=A,H'_1=B+C$. 
From $H_{\epsilon n}=H_0 + ge^{-\epsilon |t_n|}H_1$, 
we obtain 
\begin{eqnarray}
H_{\epsilon n}=H'_{\epsilon n}+(1-ge^{-\epsilon |t_n|})B. 
\label{difhep}
\end{eqnarray}
The last term $(1-ge^{-\epsilon |t_n|})B$ which is 
dependent on the method used to divide the Hamiltonian 
vanishes in the limit $g \rightarrow 1,\epsilon \rightarrow 0$, 
hence $\Omega_{\epsilon}$ remains the same. 
Thus, the density matrix in eq.(\ref{zubarev}) 
does not depend on a division of the Hamiltonian 
in the limit $g \rightarrow 1,\epsilon \rightarrow 0$. 
Therefore, 
we can conclude that the form obtained using the 
${\rm C}^{*}$ algebra 
is identical with the present form 
obtained using Keldysh formalism. 

\section{Characteristic Feature of Nonequilibrium Steady State}

The fact that $\bar{\rho}_{\epsilon}$ is the invariant 
part of the density matrix leads to the understanding 
of the dissipation in the system and also of the mechanism 
needed to reach a steady state. These are the points clarified 
by Zubarev\cite{zubarevtxt}. However, in view of its 
importance, we briefly repeat the essential parts of 
the arguments explicitly using Keldysh formalism. 

\subsection{Invariant part of density matrix and 
causality condition}

The Neumann equation in eq.(\ref{neumann}) is rewritten as 
\begin{eqnarray}
{\rm i} \frac{\partial}{\partial t} \ln \rho _{\epsilon}(t)
= [H_{\epsilon}, \ln \rho _{\epsilon}(t)]. 
\label{neumannln}
\end{eqnarray}
By using eq.(\ref{adiabatic}), $\ln \bar{\rho}_{\epsilon}$ 
can be also written as 
\begin{eqnarray}
&&
\ln \bar{\rho}_{\epsilon} =\ln \rho _{0}+
\int^{0}_{-\infty} {\rm d}t e^{-|\epsilon|t} J_{\ln \rho_{0H}}(t). 
\label{adiabaticln}
\end{eqnarray}
where $J_{\ln \rho_{0H}}(t)$ is defined 
using eq.(\ref{current}) for $A=\ln \rho_0$. 
Corresponding to eq.(\ref{adiabatic2}), 
we rewrite it as 
\begin{eqnarray}
\nonumber
&&
{\,}
\lim_{\epsilon \rightarrow 0} \ln \bar{\rho}_{\epsilon}=
\lim_{\epsilon \rightarrow 0}
\epsilon
\int^{0}_{-\infty} {\rm d}t e^{-|\epsilon|t} \ln \rho _{0H}(t) \\
&&{\qquad \quad \quad \!}=
\lim_{T_a \rightarrow \infty} 
\frac{1}{T_a}
\int^{0}_{-T_a} {\rm d}t \ln \rho _{0H}(t).
\label{zubarevdef}
\end{eqnarray}
We find that $\ln \bar{\rho}_{\epsilon}$ 
is given by 
the long-time average of the logarithm of 
evolution of the initial density matrix. 

This is nothing but Zubarev's definition 
of a nonequilibrium statistical operator\cite{zubarevtxt}. 
He has pointed out that 
the entropy production is zero  
in the local equilibrium state; 
hence, the local 
equilibrium ensembles cannot describe 
irreversible processes. 
To incorporate dissipation, 
the idea of coarse graining has been introduced. 
He has defined the nonequilibrium statistical 
operator as the long-time average 
of entropy which is 
logarithm of the local equilibrium ensemble. 
The time average has been defined  
at the negative semi-infinite interval, 
namely, the causality condition has been imposed. 
Thus, the entropy production has been 
proved to be positive, 
although the proof itself was limited in the linear response regime. 

Therefore, the adiabatic turning on of $H_1$ 
precisely corresponds to taking the 
invariant part with the causality condition. 
Therefore, we conclude that the adiabatic 
description of Keldysh formalism 
may describe irreversible processes. 
\subsection{Steady-state feature of time-correlation functions}
Based on the fact that $\bar{\rho}_{\epsilon}$ is 
the invariant part, 
we will show that the steady state is described by the 
adiabatic switching on of Keldysh formalism 
in the limit $\epsilon \rightarrow 0$. 

We start our discussion with charge current. 
Following similar calculations to obtain eq.(\ref{commutationas}), 
we show that the commutation relation between 
$\bar{\rho}_{\epsilon}$ and $U_{\epsilon}(0,t)$ 
is expressed as 
\begin{eqnarray}
&&
{\!\!\!\!\!\!\!\!\!\!\!\!\!\!\!}
[\bar{\rho}_{\epsilon},U_{\epsilon}(0,t) ]=
-\int^{0}_{t} {\rm d}t_1 J_{\bar{\rho}_{\epsilon} H}(t_1) 
U_{\epsilon}(0,t), 
\label{comutation4}
\end{eqnarray}
where $J_{\bar{\rho}_{\epsilon} H} (t_1)$ is defined by 
\begin{eqnarray}
\nonumber
&&{\!\!\!\!\!\!\!\!\!\!\!\!\!\!\!}
{\!\!\!\!\!\!}
J_{\bar{\rho}_{\epsilon} H}(t_1) 
\equiv 
- \frac{\partial}{\partial t_1} 
\bar{\rho}_{\epsilon H}(t_1) \\
&&{\;}
= U_{\epsilon}(0,t_1) {\rm i}[\bar{\rho}_{\epsilon}, 
H_{\epsilon 1}] U_{\epsilon}(t_1,0). 
\label{jrho}
\end{eqnarray}
From eq.(\ref{comutation4}), 
we rewrite the expectation value of the current as 
\begin{eqnarray}
\nonumber
&&
{\!\!\!\!\!\!}
{\rm Tr}\{ \bar{\rho}_{\epsilon} J_{cH}(t) \} 
={\rm Tr}\{ U_{\epsilon}(t,0) \bar{\rho}_{\epsilon} 
U_{\epsilon}(0,t) J_c \} \\
&&
={\rm Tr}\{ \bar{\rho}_{\epsilon} J_c \}+ 
\int^{t}_{0} {\rm d}t_1 {\rm Tr}\{ 
J_{\bar{\rho}_{\epsilon} H}(t_1) J_{cH}(t) \}. 
\label{current1}
\end{eqnarray}
We find that $J_{\bar{\rho}_{\epsilon} H}(t_1)$ 
is a key quantity for determining the 
time dependence of charge current. 
$J_{\bar{\rho}_{\epsilon} H}(t_1)$ is 
time derivative of the Heisenberg representation of 
the invariant part $\bar{\rho}_{\epsilon}$, 
which is one of the local integrals of motion 
defined by Zubarev\cite{zubarevtxt}. 

Here, we discuss $[\bar{\rho}_{\epsilon}, H_{\epsilon 1} ]$ 
in eq.(\ref{jrho}). 
By using the invariant part of $\bar{\rho}_{\epsilon}$ 
in eq.(\ref{adiabatic2}), 
the commutation relation is rewritten as 
\begin{eqnarray}
\nonumber
& &{\!\!\!\!\!}[\bar{\rho}_{\epsilon}, H_{\epsilon 1}] 
= -{\rm i} \epsilon \int^{0}_{-\infty} {\rm d}t_2 
e^{-\epsilon |t_2|} J_{\rho_0 H}(t_2) \\
& & {\qquad \qquad \quad}
+ \epsilon \int^{0}_{-\infty} {\rm d}t_2 
g_{\epsilon} [ \rho_{0H}(t_2), H_{\epsilon 1} ], 
\label{comutation5}
\end{eqnarray}
where we used 
\begin{eqnarray}
\nonumber
&&{\!\!\!\!\!\!\!\!\!\!\!\!\!\!\!\!\!\!\!\!\!}
\left[ \rho_{0H}(t_2), H_{\epsilon 1} \right] \\
\nonumber
&&{\!\!\!\!\!\!\!\!\!\!\!\!\!\!\!}
= [\rho_{0H}(t_2), H_{\epsilon 2}]
+[\rho_{0H}(t_2), H_{\epsilon 1}-H_{\epsilon 2}] \\
&&{\!\!\!\!\!\!\!\!\!\!\!\!\!\!\!}
= -{\rm i}J_{\rho_0 H}(t_2)
+g(e^{-\epsilon|t_1|}-e^{-\epsilon|t_2|}) 
[\rho_{0H}(t_2), H_{1}] 
\label{comutation6}
\end{eqnarray}
and $g_{\epsilon} \equiv g e^{-\epsilon|t_2|}
(e^{-\epsilon|t_1|}-e^{-\epsilon|t_2|})$. 
From eq.(\ref{adiabatic}), the first term is 
given by 
\begin{eqnarray}
-{\rm i} \epsilon 
\int^{0}_{-\infty} {\!\!\!} {\rm d}t_2 
e^{-\epsilon |t_2|} J_{\rho_0 H}(t_2) 
= -{\rm i} \epsilon (\bar{\rho}_{\epsilon}  -\rho_0 ).
\label{firstterm}
\end{eqnarray}
Due to $g_{\epsilon}$, 
the domain of the integration of the second term 
in eq.(\ref{comutation5}) vanishes in the limit 
$\epsilon \rightarrow 0$; hence, we can omit 
the second term. Thus, the commutation relation 
in the limit $\epsilon \rightarrow 0$ becomes 
\begin{eqnarray}
\lim_{\epsilon \rightarrow 0}
[\bar{\rho}_{\epsilon}, H_{\epsilon 1}] 
= 
\lim_{\epsilon \rightarrow 0}
-{\rm i} \epsilon (\bar{\rho}_{\epsilon}-\rho_0). 
\label{comutation7}
\end{eqnarray}

On the basis of the commutation relation, the second term of 
eq.(\ref{current1}) is shown to be 
proportional to $\epsilon$, expressed as 
\begin{eqnarray}
\nonumber
&&{\!\!\!\!\!\!\!\!\!\!\!\!\!\!}
\lim_{\epsilon \rightarrow 0}
{\rm Tr}\{ J_{\bar{\rho}_{\epsilon} H}(t_1) J_{cH}(t) \} 
= \lim_{\epsilon \rightarrow 0}
{\rm Tr}\{ {\rm i} [\bar{\rho}_{\epsilon}, H_{\epsilon 1}] 
J'_{cH}(t_1,t) \} \\
&&{\qquad \qquad \qquad \qquad }
= \lim_{\epsilon \rightarrow 0}
\epsilon \cdot {\rm Tr}\{ (\bar{\rho}_{\epsilon}-\rho _{0}) 
J'_{cH} (t_1,t) \} \nonumber \\
&&{\qquad \qquad \qquad \qquad }
= 0, 
%
\label{current2}
\end{eqnarray}
where 
$J'_{cH} (t_1,t) \equiv 
U_{\epsilon}(t_1,0) J_{cH}(t) U_{\epsilon}(0,t_1)$. 
The expectation values of any operator 
for $\bar{\rho}_{\epsilon}$ and $\rho _{0}$ 
are assumed to exist in the limit  
$\epsilon \rightarrow 0$ in Keldysh formalism. 
Therefore, 
$\epsilon \cdot {\rm Tr} \{ \cdots \}$ vanishes 
in the limit $\epsilon \rightarrow 0$. 
In other words, $\bar{\rho}_{\epsilon}$ and $H_{\epsilon 1}$ 
inside the trace commute with each other in the limit 
$\epsilon \rightarrow 0$. 
We have shown that 
hence, the second term of eq. (\ref{current1}) vanishes. 

Thus, we conclude that 
the expectation value of the current becomes 
independent of time in the limit $\epsilon \rightarrow 0$. 
\begin{eqnarray}
{\!\!\!\!}
\lim _{\epsilon \rightarrow 0} \langle J_{cH}(t) \rangle=
\lim_{\epsilon \rightarrow 0} \langle J_c \rangle . 
\label{chrgecurrent2}
\end{eqnarray}
Following the same type of analysis, the time-dependent 
correlation function of charge current is obtained as 
\begin{eqnarray}
{\qquad \ \>}
\lim _{\epsilon \rightarrow 0} 
\langle J_{cH}(t) J_{cH}(t^{'}) \rangle
=\lim _{\epsilon \rightarrow 0} 
\langle J_{cH}(t-t^{'}) J_{cH}(0) \rangle. 
\label{chrgecurrent22}
\end{eqnarray}

The above two equations prove that the steady state is 
actually reached by the adiabatic switching-on of the 
perturbation term in Keldysh formalism. 

\section{Nonequilibrium Kubo Formula}

The charge current and differential conductance 
are respectively defined by 
\begin{eqnarray}
J (V) \equiv 
\lim _{g \rightarrow 1, \epsilon \rightarrow 0} 
\langle J_{cH}(t) \rangle = 
\lim _{g \rightarrow 1, \epsilon \rightarrow 0} 
\langle J_c \rangle, 
\label{recurrent}
\end{eqnarray}
\begin{eqnarray}
G(V) \equiv \frac{\partial J (V)}{\partial V}. 
\label{conductance}
\end{eqnarray}
In the last equality of eq.(\ref{recurrent}), 
the steady-state feature of eq.(\ref{chrgecurrent2}) is used. 

The differential conductance is 
obtained by differentiating 
the MacLennan-Zubarev form of the density matrix 
$\bar{\rho}_{\epsilon}$, eq.(\ref{zubarev}), 
with respect to $V$. 
\begin{eqnarray}
\nonumber
& &G(V)=\lim _{g \rightarrow 1, \epsilon \rightarrow 0}
\beta {\rm Tr}  \biggl\{  
J_c {\bigg(} e/2(N_L -N_R)  \\
& &{\qquad \qquad \qquad} \left. \left. 
+ \int^{0}_{-\infty} {\!\!\!} {\rm d}t e^{-\epsilon |t|} 
J_{cH}(t) + \frac{\partial \Omega _0}{\partial V}
\right) \bar{\rho}_{\epsilon}  \right\}. 
\nonumber \\
\label{gv1}
\end{eqnarray}

By substituting the identity 
$S_{\epsilon}(0,-\infty) S_{\epsilon}(-\infty,0) \equiv 1$, 
$\partial \Omega _0 /\partial V$ is calculated as 
\begin{eqnarray}
\nonumber
&&{\!\!\!\!\!\!\!\!\!\!\!\!\!\!}
\frac{\partial \Omega _0}{\partial V}
=-{\rm Tr}\{ e/2 (N_L -N_R) \rho _{0} \} \\
\nonumber
&&
=-{\rm Tr}\{ S_{\epsilon}(0,-\infty) e/2 (N_L -N_R) 
S_{\epsilon}(-\infty,0)\rho _{0} \} \\
&&
=- e/2 \langle N_L -N_R \rangle  - \int^{0}_{-\infty} 
{\!\!\!} {\rm d}t e^{-\epsilon |t|} \langle J_{cH}(t) \rangle ,
\label{difomega0}
\end{eqnarray}
where in the last step eq.(\ref{adiabatic}) is used.
Thus, we obtain the expression 
\begin{eqnarray}
\nonumber
&&G(V)=\lim _{g \rightarrow 1, \epsilon \rightarrow 0}
\beta 
{\bigg(}
\langle \delta J_c e/2(\delta N_L -\delta N_R) \rangle \\
&& \left. {\qquad \qquad \qquad}
+\int^{0}_{-\infty} {\!\!\!} {\rm d}t e^{-\epsilon |t|} 
\langle\delta J_{cH}(0) \delta J_{cH}(t)\rangle   \right), 
\nonumber \\
\label{gv2}
\end{eqnarray}
where 
$\delta A \equiv A -\langle A \rangle $ and 
$\langle A\delta B\rangle  =\langle \delta A \delta B\rangle $ 
are used. 

Now, we rewrite the differential conductance into 
another expression. In preparation we derive 
several commutation relations. 
On the basis of 
$[H_0,e/2(N_L -N_R)]=0$, 
\begin{eqnarray}
S_{\epsilon}(0,-\infty) [H_0,e/2(N_L -N_R)] 
S_{\epsilon}(-\infty,0) = 0 
\label{comutation2.1}
\end{eqnarray}
is obtained, which leads to 
\begin{eqnarray}
\nonumber
&&{\biggl[} {\,} H_0 + \int ^{0}_{-\infty} 
{\!\!\!} {\rm d}t e^{-\epsilon |t|} {\rm d}t J_{eH}(t), \\
&& {\qquad} e/2(N_L -N_R) + \int ^{0}_{-\infty}
{\!\!\!} {\rm d}t e^{-\epsilon |t|} {\rm d}t J_{cH}(t)\ {\bigg]} =0,
\nonumber \\
\label{comutation2}
\end{eqnarray}
From this equation, one can derive 
\begin{eqnarray}
\nonumber
&&\left[ {\,} \bar{\rho}_{\epsilon}, H_0 + \int ^{0}_{-\infty}
{\!\!\!} {\rm d}t e^{-\epsilon |t|} {\rm d}t J_{eH}(t) \right] =0, \\
&&\left[ {\,} \bar{\rho}_{\epsilon},
e/2(N_L -N_R) + \int ^{0}_{-\infty}
{\!\!\!} {\rm d}t e^{-\epsilon |t|} {\rm d}t J_{cH}(t)\ \right] =0. 
\nonumber \\
\label{comutation3}
\end{eqnarray}
Using the second commutation relation of eq.(\ref{comutation3}) 
for the correlation functions in eq.(\ref{gv1}), 
we can exchange current and the invariant part 
of $e/2(N_L-N_R)$, which leads to 
\begin{eqnarray}
\nonumber
&&G(V)=\lim _{g \rightarrow 1, \epsilon \rightarrow 0}
\beta 
{\bigg(} 
\langle e/2(\delta N_L -\delta N_R) \delta J_c  \rangle  \\
&&\left. {\qquad \qquad \qquad}
+\int^{0}_{-\infty} {\!\!\!} {\rm d}t e^{-\epsilon |t|} 
\langle\delta J_{cH}(t) \delta J_{cH}(0) \rangle   \right). 
\nonumber \\
\label{gv3}
\end{eqnarray}

Moreover in eqs. (\ref{gv2}) and (\ref{gv3}) 
the time integrals can be rewritten at the 
positive semi-infinite interval 
with the steady-state property of eq.(\ref{chrgecurrent2}). 
Thus, four distinct expressions of $G(V)$ are 
obtained. Their sum, divided by four, 
gives the symmetrized expression of differential 
conductance, 
\begin{eqnarray}
\nonumber
&&G(V)=\lim _{g \rightarrow 1, \epsilon \rightarrow 0}
\frac{\beta}{4}
\langle \{ \delta J_c , e(\delta N_L -\delta N_R) \}\rangle  \\
&&{\qquad \ }
+\lim _{g \rightarrow 1, \epsilon \rightarrow 0}
\frac{\beta}{4}
\int^{\infty}_{-\infty} {\!\!\!} {\rm d}t e^{-\epsilon |t|} 
\langle \{ \delta J_{cH}(t), \delta J_{cH}(0) \} \rangle 
\nonumber \\
\label{symmeticgv}
\end{eqnarray}
where $\{A,B\}=AB+BA$ represents the anticommutation relation.

We find that 
differential conductance is determined by two contributions: 
the current-current correlation 
function and an unusual correlation function between 
current and the difference of the two particle 
numbers of reservoirs. 
This result directly reflects 
the fact that 
the conjugate of the bias $V$ 
is the invariant part of $e/2(N_L -N_R)$ 
in MacLennan-Zubarev form of the density matrix. 
Equation (\ref{symmeticgv}) is the main result of this study 
and may be considered as the generalization of the Kubo formula 
in a nonequilibrium situation for a general system 
of a finite conductor connected to the left and right reservoirs. 
\section{Physical Quantities}
\subsection{Shot noise}

We introduce noise power as using 
\begin{eqnarray}
{\quad \,\,}S(V) \equiv 
\lim _{g \rightarrow 1, \epsilon \rightarrow 0}
\int^{\infty}_{-\infty} {\!\!\!} {\rm d}t e^{-\epsilon |t|} 
\langle \{ \delta  J_{cH}(t), \delta J_{cH}(0) \} \rangle 
{\!\!}
\label{noisepower}
\end{eqnarray}
and define 
the current-charge correlation as 
\begin{eqnarray}
{\!\!\!}
S_{\rm sh}(V) \equiv 
- \lim _{g \rightarrow 1, \epsilon \rightarrow 0}
\langle \{ \delta J_c , e(\delta N_L -\delta N_R) \}\rangle .
\label{sh}
\end{eqnarray}
By using $S(V)$ and $S_{\rm sh}(V)$, 
the differential conductance is given by 
\begin{eqnarray}
G(V)=\frac{\beta}{4}S(V) - \frac{\beta}{4} S_{\rm sh}(V).
\label{symmetricgv2}
\end{eqnarray}
Differential conductance and noise power 
are well-known observable quantities in mesoscopic systems. 
We propose that 
the $S_{\rm sh}(V)$ given by eq.(\ref{sh}) 
be defined as shot noise in general, 
for both interacting and non-interacting systems. 

To confirm that eq.(\ref{sh}) is actually shot noise, 
we discuss 
the simplest example of a noninteracting 
Anderson model for a dot attached to 
the left and right reservoirs, 
\begin{eqnarray}
\nonumber
{\ }
H=\sum_{pk \sigma} c^{\dagger}_{p k \sigma} 
c_{pk \sigma} + \varepsilon_d \sum_{\sigma} n_{\sigma} 
+ \sum_{pk\sigma} 
V_{pk \sigma} c^{\dagger}_{p k \sigma} d_{\sigma} + h.c.,
{\!\!\!\!\!\!\!\!\!\!} \\
\label{u0anderson}
\end{eqnarray}
where $c_{pk \sigma}$ is the annihilation operator 
of the electron with the spin $\sigma$
in the left ($p$=$L$) or right ($p$=$R$) reservoir. 
$d_{\sigma}$ is 
the annihilation operator at the dot and 
$n_{\sigma}=d^{\dagger}_{\sigma} d_{\sigma}$. 
$V_{pk \sigma}$ is the hybridization
between the dot and leads with $p=L,R$. 

It is easy to calculate charge current 
using the Keldysh Green function 
\begin{eqnarray}
J(V)=2e \int^{\infty}_{-\infty} 
\frac{{\rm d} \omega}{2\pi} T (\omega) (f_L -f_R), 
\label{u0current}
\end{eqnarray}
where 
$f_{L,R} =1/(1+e^{\beta (\omega \mp eV/2)})$ 
are the Fermi distribution functions. 
Transmission probability is defined as 
\begin{eqnarray}
T (\omega) = \frac{\Gamma_L \Gamma_R}
{\Gamma_L + \Gamma_R} 
\frac{2 \Gamma}{(\omega -\varepsilon_d)^2+\Gamma^2}, 
\label{transmission}
\end{eqnarray}
where $\Gamma_{L,R} =\sum_k 
2\pi |V_{L,Rk\sigma}| 
\delta (\omega -\varepsilon_{L,Rk})$ 
and $\Gamma =(\Gamma_L +\Gamma_R)/2$. 
The differential conductance is obtained 
by differentiating $J(V)$ with respect $V$ as using 
\begin{eqnarray}
\nonumber
& &{\!\!\!\!\!\!\!}
G(V)=\beta e^2
\int^{\infty}_{-\infty}
\frac{{\rm d} \omega}{2\pi} T (\omega) 
(f_L(1-f_L)+f_R(1-f_R)). 
\nonumber \\
\label{u0gv}
\end{eqnarray}

On the basis of the definition of eq. (\ref{noisepower}), 
the noise power $S(V)$ may be written as 
\begin{eqnarray}
\nonumber
& &{\!\!\!\!\!\!\!\!\!}
S(V)= 4 e^2 
\int^{\infty}_{-\infty}
\frac{{\rm d} \omega}{2\pi} T (\omega) 
(f_L (1-f_L)+f_R(1-f_R)) \\
& & {\quad \!} +
4 e^2 
\int^{\infty}_{-\infty}
\frac{{\rm d} \omega}{2\pi} T (\omega) (1-T (\omega))
(f_L-f_R)^2. 
\label{u0noisepower}
\end{eqnarray}
In these way, we have confirmed that noise power 
is given by the sum of the equilibrium noise, 
the first line of eq.(\ref{u0noisepower}), 
and shot noise. This result is well known 
in noninteracting systems \cite{beenakker,blanter}. 

On the other hand, one can show that 
$S_{\rm sh}(V)$ is given by 
\begin{eqnarray}
&&{\!\!\!\!\!\!\!\!\!\!\!\!\!\!}
S_{\rm sh}(V)=4 e^2 
\int^{\infty}_{-\infty}
\frac{{\rm d} \omega}{2\pi} T (\omega) 
(1-T (\omega))
(f_L-f_R)^2
\label{u0noisepower2}
\end{eqnarray}
for a noninteracting system. 
This is nothing but the second term of 
eq.(\ref{u0noisepower}). 

Therefore for the noninteracting system 
discussed here, 
we confirmed that $S_{\rm sh}(V)$ expresses 
shot noise and that eq.(\ref{symmetricgv2}) 
actually holds. 
\subsection{Generalization of Nyquist-Johnson relation}

In eq.(\ref{symmetricgv2}), we have found that 
conductance is given by 
the difference between noise power and shot noise. 
We would like to 
stress again that this general expression is 
also valid in interacting systems, and not limited 
to the noninteracting systems discussed 
in the literature so far. 
Therefore, we propose eq.(\ref{symmetricgv2}) 
as a {\it nonequilibrium{\ }identity} between 
physical quantities, which is 
a generalization of the Nyquist-Johnson relation 
into many-body systems in a nonequilibrium case. 
Below, we show that 
eq.(\ref{symmetricgv2}) is actually 
a generalization of the Nyquist-Johnson relation. 

Let us first discuss the linear response regime. 
The density matrix at $V=0$ can be obtained 
using eq.(\ref{gellman}). 
It is well known that 
the last term of eq.(\ref{gellman}) describes 
the sum of all vacuum loops, 
effectively giving the correction of the thermodynamic 
potential $\Delta \Omega \equiv \Omega - \Omega_0$ 
\cite{Keldysh}; hence, the equilibrium ensemble 
becomes 
\begin{eqnarray}
\lim_{g \rightarrow 1, \epsilon \rightarrow 0}
\bar{\rho}_{\epsilon} |_{V=0} 
= e^{-\beta (H-\Omega)} . 
\label{eqdensity}
\end{eqnarray}
In any equilibrium ensemble, 
we can calculate the trace with the eigenfunctions of 
$H |n \rangle = E_n |n \rangle $ 
and $J_c ={\rm i}[e(N_L -N_R),H]$
so that 
\begin{eqnarray}
S_{\rm sh}(0)=-{\rm i}e^2 \sum_{n=0}^{\infty} 
\langle n | [(\delta N_L -\delta N_R)^2 ,E_n] |n \rangle 
=0, 
\label{jhonson}
\end{eqnarray}
is derived in general. 
$S_{\rm sh}(0)$ vanishes in the linear response regime. 
Thus, we obtain 
\begin{eqnarray}
4 k_{\rm B} T G(0) = S(0).
\label{jhonson}
\end{eqnarray}
This corresponds to a Nyquist-Johnson relation 
in the linear response regime 
even for systems with electron correlation 
\cite{beenakker,blanter}. 

This relation can also be derived using the fluctuation 
dissipation theorem. 
Thus the nonequilibrium identity 
eq.(\ref{symmetricgv2}) shows 
that a naive generalization of $4 k_{\rm B} T G(V) = S(V)$
in the nonequilibrium case is not possible 
and there is the additional contribution $S_{\rm sh}(V)$, 
which plays an essential role 
in a nonequilibrium situation. 

Finally, let us discuss 
how to obtain 
shot noise experimentally. 
As we have discussed in the preceding subsection 
for a noninteracting system, noise power 
may be divided into equilibrium noise 
and shot noise. 
Experimentally, we measure 
only the noise power. In actual cases, 
it has been discussed that noise power 
for $eV>k_{\rm B} T_K$ includes 
shot noise significantly. 
On the other hand, the nonequilibrium identity 
suggests that shot noise as the difference 
between noise power and differential conductance 
expressed as 
\begin{eqnarray}
S_{sh}(V)= S(V)- 4 k_{\rm B} T G(V), 
\label{shotnoise3}
\end{eqnarray}
both of which are measurable quantities. 
Thus, we may use this identity as the 
definition of shot noise for both  
interacting and noninteracting systems. 
\section{Conclusion}

We have clarified that 
the density matrix obtained by the Keldysh formalism 
of a finite conductor in series with reservoirs 
is an example of MacLennan-Zubarev ensembles. 
On the basis of this fact, we pointed out that 
the adiabatic switching-on of the perturbative term 
of Keldysh formalism corresponds to 
taking the invariant part with the causality condition 
introduced by Zubarev. 
From this fact, we conclude that 
Keldysh formalism 
may describe an irreversible process and 
the steady-state feature of the time-correlation functions 
For proof of this 
steady-state feature, 
it is essential 
to understand the fact that the density matrix and 
total Hamiltonian commute in the expectation values. 

Using the MacLennan-Zubarev form, 
we have derived the general expression 
of differential conductance, 
which is determined by 
the current-current correlation 
function and the correlation 
function between 
current and the difference of the two particle 
numbers of reservoirs. 
We call it the nonequilibrium Kubo 
formula of a finite conductor attached to reservoirs. 
It is well known that noise power is given by 
the current-current correlation function. 
We have proposed that 
the correlation function between 
current and the difference of the two particle 
numbers of reservoirs gives 
the shot noise of the system, in general. 
This fact is confirmed by explicit calculations 
using the noninteracting Anderson model. Therefore, 
we propose that the nonequilibrium identity 
between differential conductance, noise power 
and shot noise is 
a generalization of the Nyquist-Johnson relation 
in general. 
\section*{Acknowledgments}
The author would like to thank 
K. Ueda, T. Kato and K. Saito for valuable discussions. 
This research was partially supported by 
the Ministry of Education, Science, Sports 
and Culture, Grant-in-Aid for Young Scientists (B), 2005, 
17740187. 

\begin{thebibliography}{99} 
%
\bibitem{Keldysh}
L. V. Keldysh: 
Sov. Phys. JETP. {\bf 20} (1965) 1018.
%
\bibitem{langreth}
D. C. Langreth: 
in {\it Linear and Nonlinear Transport in Solids}, 
eds. J. T. Devreese and V. E. Van Doren 
(Plenum Press, New York, 1976) 
Vol. 17 of NATO ASI, Series B: Physics.
%
\bibitem{rammer}
J. Rammer: 
Rev. Mod. Phys. {\bf 58} (1986) 323.
%
\bibitem{kamenev1}
A. Kamenev: 
in {\it Strongly Correlated Fermions and Bosons in 
Low-Dimensional Disordered Systems}, 
eds. I. V. Lerner, B. L. Althsular, V. I. Fal'ko, T. Giamarchi 
(Kluwer Academic Publishers, Dordrecht, 2002) 
Nato Science Series; Sub-series II, Mathematics, 
Physics and Chemistry, Vol 72.
%
\bibitem{kamenev2}
A. Kamenev: 
in Lectures notes for 2004 Les Houches Summer School on 
"Nanoscopic Quantum Physics". 
%
\bibitem{Hershfield2}
S. Hershfield: 
Phys. Rev. Lett. {\bf 70} (1993) 2134. 
%
\bibitem{Hershfield3}
A. Schiller and S. Hershfield: 
Phys. Rev. B {\bf 51} (1995) 12896; 
A. Schiller and S. Hershfield: 
Phys. Rev. Lett. {\bf 77} (1996) 1821; 
A. Schiller and S. Hershfield: 
Phys. Rev. B {\bf 58} (1998) 14978.  
%
%
\bibitem{stasaki1}
S. Tasaki: 
Chaos, Solitons and Fractals {\bf 12} (2001) 2657; 
in {\it Statistical Physics}, 
ed. M. Tokuyama and H. E. Stanley 
(AIP Press, New York, 2000) p. 356; 
in {\it Quantum Information III}, 
ed. T. Hida and K. Saito 
(World Scientific, Singapore, 2001) p. 157.
%
%
\bibitem{stasaki2}
S. Tasaki and T. Matsui: 
in {\it Fundamental Aspects of Quantum Physics}, 
eds. L. Accardi and S. Tasaki (World Scientific, 2003) p. 100.
%
%
\bibitem{stasaki3}
J. Takahashi and S. Tasaki: 
J. Phy. Soc. Jap. {\bf 75} (2006) 94712.
%
%
\bibitem{Caroli}
C. Caroli, R. Combescot, P. Nozi$\grave{e}$res, and D. S-James: 
J. Phys. C: Solid St. Phys. {\bf 4} (1971) 916.
%
%
\bibitem{Hershfield}
S. Hershfield, J. H. Davies, and J. W. Wilkins: 
Phys. Rev. Lett. {\bf 67} (1991) 3720; 
Phys. Rev. B. {\bf 46} (1992) 7046.
%
\bibitem{Meir}
Y. Meir, N. S. Wingreen, and P. A. Lee: 
Phys. Rev. Lett. {\bf 70} (1993) 2601;
N. S. Wingreen and Y. Meir: 
Phys. Rev. B. {\bf 49} (1994) 11040. 
%
\bibitem{nozierestxt}
P. Nozi$\grave{e}$res: 
{\it Theory of Interacting Fermi Systems} 
(Benjamin New York, 1964). 
%
\bibitem{fetter}
See, for example, A. L Fetter and J. D. Walecka: 
{\it Quantum Theory of Many Particles} 
(McGraw-Hill, New York, 1971). 
%
\bibitem{zubarevtxt}
D. N. Zubarev: 
{\it Nonequilibrium Statistical Thermodynamics} 
(Consultants, New York, 1974). 
%
\bibitem{maclennan}
J. A. MacLennan, Jr.: 
Adv. Chem. Phys. {\bf 5} (1963) 261.
%
%
\bibitem{beenakker}
M. J. M. de Jong and C. W. J. Beenakker: 
in {\it Mesoscopic Electron Transport},
eds. Lydia L. Sohn, Leo P. Kouwenhoven, and Gerd Schon 
(Kluwer Academic Publishers, Dordrecht, 1997)
Vol. 345 of NATO ASI, Sereies E: Applied Sciences, p. 225. 
%
%
%
%
\bibitem{blanter}
See, for example, Ya. M. Blanter and M. B$\ddot{u}$ttiker: 
Physics Reports {\bf 336} (2000) p. 1.
%
\end{thebibliography}
\end{document}